\documentclass[12pt]{article}
\usepackage{graphicx}
\def\beq{\begin{equation}}
\def\eeq{\end{equation}}
\def\bea{\begin{eqnarray}}
\def\eea{\end{eqnarray}}
\def\nn{\nonumber}
\def\roughly#1{\mathrel{\raise.3ex\hbox
{$#1$\kern-.75em\lower1ex\hbox{$\sim$}}}}

\def\bra#1{\left\langle #1\right|}
\def\ket#1{\left| #1\right\rangle}
\def\bd{B_d^0}
\def\bdbar{{\bar B}^0_d}
\def\bs{B_s^0}
\def\bsbar{{\bar B}^0_s}
\def\barp{{\raise.35ex\hbox{${\sss (}$}}---{\raise.35ex\hbox{${\sss )}$}}}
\def\bdbarp{\hbox{$B_d$\kern-1.4em\raise1.4ex\hbox{\barp}}}
\def\sss{\scriptscriptstyle}
\def\ks{K_{\sss S}}
\def\kbar{{\bar K}^0}
\def\kstarbar{{\bar K}^*}
\def\barpk{{\raise.35ex\hbox
{${\sss (}$}}--{\raise.35ex\hbox{${\sss )}$}}}
\def\kbarp{\hbox{$K$\kern-0.9em\raise1.4ex\hbox{\barpk}}}
\def\kstarbar{{\bar K}^*}
\def\Puc{{\cal P}_{uc}}
\def\Ptc{{\cal P}_{tc}}
\def\tildePuc{{\widetilde{\cal P}}_{uc}}
\def\tildePtc{{\widetilde{\cal P}}_{tc}}
\def\tildeDelta{\widetilde{\Delta}}
\begin{document}

\begin{flushright}  
UdeM-GPP-TH-03-115 \\
McGill 03/24 \\
\end{flushright}

\begin{center}
\bigskip
{\Large \bf CP Violation Beyond the Standard Model\footnote{Talk given
at the {\it $9^{th}$ International Conference on B-Physics at Hadron
Machines --- BEAUTY 2003}, Carnegie Mellon University, Pittsburgh, PA,
USA, October 2003.}}  \\

\bigskip
\bigskip
{\large David London\footnote{london@lps.umontreal.ca}}
\end{center}

\begin{center}
{\it Physics Department, McGill University, \\ 3600 University St.,
Montr\'eal QC, Canada H3A 2T8}\\
{\rm and}\\
{\it Laboratoire Ren\'e J.-A. L\'evesque, Universit\'e de Montr\'eal, \\
C.P. 6128, succ. centre-ville, Montr\'eal, QC, Canada H3C 3J7}
\end{center}

\begin{center} 
\bigskip (\today)
\vskip0.5cm
{\Large Abstract\\}
\vskip3truemm
\parbox[t]{\textwidth} { I review CP-violating signals of physics
  beyond the standard model in the $B$ system. I examine the prospects
  for finding these effects at future colliders, with an emphasis on
  hadron machines.}
\end{center}

\thispagestyle{empty}
\newpage
\setcounter{page}{1}
% Decrease texheight (for preprint numbers) again
%\textheight 23.0 true cm
\baselineskip=14pt

\section{Introduction}

I have been asked to review the various CP-violating signals for
physics beyond the standard model (SM) in the $B$ system, with a
particular emphasis on future hadron colliders. Now, in any discussion
of this type, one has to consider the following question: will this
new physics (NP) be discovered directly or not? If the assumption is
that the NP will not be observed directly at hadron colliders, then
the aim of measuring CP violation in the $B$ system is to find
evidence for NP. This ``discovery signal'' study is model-independent.
That is, if some signal is seen, one will know that NP is present, but
one will not know what kind of NP it is. On the other hand, if one
assumes that this NP can be produced directly at hadron colliders,
then its discovery will also probably reveal its identity, though not
the details of its properties. In this case, the study of $B$ physics
is still useful -- it will furnish ``diagnostic tests'' of this NP.
The point is that the future study of CP violation in the $B$ system
is important, though what we will learn depends on what is discovered
(or not) through other measurements. In particular, it is essential to
consider both possibilities --- that the NP is discovered directly, or
not --- in any discussion of CP violation signals of NP at future
colliders \cite{Cahn}. In this talk I will attempt to address both of
these scenarios.

If new physics exists, it can affect the $B$ system in many different
ways:
\begin{enumerate}

\item it can lead to new effects in $\bs$--$\bsbar$ mixing or the
$b\to s$ penguin amplitude, i.e.\ in the $b\to s$ flavour-changing
neutral current (FCNC),

\item it can enter the $b\to d$ FCNC, i.e.\ $\bd$--$\bdbar$ mixing or
the $b\to d$ penguin,

\item it can affect tree-level decays, such as $b \to c {\bar q} q'$,
$b \to u {\bar q} q'$, though this is less favoured theoretically.

\end{enumerate}
Of course, any particular NP model may contain all of these effects,
and all three classes of signals should be considered.

\section{A Sign of New Physics?}

As is well known, there is a hint of a discrepancy in CP violation in
$\bd(t) \to \phi \ks$ -- the Belle measurement of $\beta$ from this
mode disagrees with that obtained from $\bd(t) \to J/\psi \ks$, though
there is no disagreement in the BaBar measurement \cite{Grossman}. If
this discrepancy is confirmed, it would point to new physics in the
${\bar b} \to {\bar s} s {\bar s}$ penguin amplitude, i.e. in the $b
\to s$ FCNC. Many models of NP have been proposed to explain this
effect: $Z$- or $Z'$-mediated FCNC's, nonminimal supersymmetry (SUSY),
SUSY with R-parity violation, left-right symmetric models, anomalous
$t$-quark couplings, etc.\ \cite{bsssmodels}. If this effect is
confirmed, we will want to distinguish among these models, either
through other $B$-physics measurements, or through direct searches at
hadron colliders.

This measurement raises an interesting question: is only the ${\bar b}
\to {\bar s} s {\bar s}$ decay affected, or are all $b\to s$ FCNC
amplitudes affected? For example, is there sizeable NP in
$\bs$--$\bsbar$ mixing? This question can be answered by making
measurements of a variety of $B$ decays.

One key task of hadron colliders is the measurement of $\bs$--$\bsbar$
mixing. This is of great interest in any case, but the potential
discrepancy in $\bd(t) \to \phi \ks$ only serves to emphasize its
importance. In order to make this measurement, it will be necessary to
resolve oscillations in the $B_s$ system. Once it has been
demonstrated that this is possible, one can turn to CP tests involving
$\bs$ mesons.

Even if new physics is discovered directly, one cannot test the CP
nature of the NP couplings to ordinary particles -- this is the domain
of $B$ physics. Hadron colliders will make several important CP
measurements involving $\bs$ mesons:
\begin{itemize}

\item Indirect CP violation in $\bs(t) \to D_s^+ D_s^-$, $J/\psi
  \phi$, $J/\psi \eta'$, etc. This measures the phase of
  $\bs$--$\bsbar$ mixing, which is $\simeq 0$ in the SM.

\item The measurement of ${\cal A}_{CP}^{mix}(\bs(t) \to D_s^\pm
  K^\mp)$ probes $\gamma$ in the SM \cite{BsDsK}. In fact, this might
  possibly be the first direct measurement of this CP phase. Its value
  can be compared to that obtained from ${\cal A}_{CP}(B^\pm \to D
  K^\pm)$ at $B$-factories \cite{BDK}.

\item Mixing-induced CP asymmetry in ${\cal A}_{CP}^{mix}(\bs(t) \to
  \phi\phi)$. This decay is analogous to $\bd(t) \to \phi\ks$. Here,
  one will need to perform an angular analysis, discussed in more
  detail below. Within the SM, this CP asymmetry is expected to be
  $\simeq 0$.

\end{itemize}
In all cases, any discrepancy with the SM prediction points
specifically to new physics, with new phases, in $\bs$--$\bsbar$
mixing and/or the $b\to s$ penguin. (Note that not all models of NP
predict new phases. For example, in the minimal supersymmetric SM with
minimal flavour violation, there are no new phases --- the couplings
of all SUSY particles track the CKM matrix.)

As an aside, suppose that the phase of $\bs$--$\bsbar$ mixing is
measured in, say, ${\cal A}_{CP}^{mix}(\bs(t) \to \Psi \phi)$. The CKM
phase $\chi \sim$ 2-5\% is extracted. Within the SM \cite{sinchi},
\beq
\sin\chi = \left\vert \frac{V_{us}}{V_{ud}} \right\vert^2
\frac{\sin\beta \sin(\gamma-\chi)}{\sin(\beta+\gamma)} ~.
\eeq
A discrepancy in this relation points to the presence of NP, though we
can't pinpoint precisely where it enters.

\section{Direct CP Violation}

Other good tests for new physics involve direct CP asymmetries. Like
any CP-violating signal, direct CP violation can only come about when
there are two interfering amplitues. In this case, such CP violation
corresponds to a difference in the rate for a $B$ decay process and
its CP-conjugate. If a given decay has only a single amplitude, then
the direct CP asymmetry must vanish.

There are many decays which are dominated by a single amplitude in the
SM.  Examples of these include $B \to J/\psi K$ and $\phi K$, $\bd \to
D_s^+ D^-$, $\bs \to D_s^+ D_s^-$, $B_c^+ \to J/\psi \pi^+$, etc. If a
direct CP asymmetry is measured in any of these modes, it implies the
presence of NP in a penguin or tree amplitude. Note that many models
of NP affect $b\to s$ or $b\to d$ penguin amplitudes; fewer affect
tree amplitudes. A complete study of direct CP asymmetries will probe
various NP models. If NP has already been found, this is a good way to
study the new couplings.

One particularly useful decay is $B^+ \to \pi^+ K^0$. In the SM, we
have $|A(B^+ \to \pi^+ K^0)| \simeq |A(B^- \to \pi^- {\bar K}^0)|$.
Thus, any direct CP violation implies new physics, specifically in the
$b\to s$ penguin. In this case the transition ${\bar b} \to {\bar s} d
{\bar d}$ is affected. (Note that there is also a hint of NP in $B\to
K\pi$ \cite{Grossman}. This is a good way of testing for this NP.)

\section{Triple Products}

One potential weakness of direct CP asymmetries is that
\beq
{\cal A}_{CP}^{dir} \propto \sin\phi \sin\delta ~,
\label{directCPV}
\eeq
where $\phi$ and $\delta$ are, respectively, the weak and strong phase
differences between the SM and NP amplitudes. Thus, if $\delta = 0$,
${\cal A}_{CP}^{dir} = 0$, even if there is a NP contribution. This
possibility can be addressed by measuring in addition triple-product
correlations (TP's).

Triple product correlations take the form $\vec v_1 \cdot (\vec v_2
\times \vec v_3)$, where each $v_i$ is a spin or momentum. TP's are
odd under time reversal (T) and hence, by the CPT theorem, also
constitute potential signals of CP violation. One can establish the
presence of a nonzero TP by measuring a nonzero value of the asymmetry
\beq
A_{\sss T} \equiv 
{{\Gamma (\vec v_1 \cdot (\vec v_2 \times \vec v_3)>0) - 
\Gamma (\vec v_1 \cdot (\vec v_2 \times \vec v_3)<0)} \over 
{\Gamma (\vec v_1 \cdot (\vec v_2 \times \vec v_3)>0) + 
\Gamma (\vec v_1 \cdot (\vec v_2 \times \vec v_3)<0)}} ~,
\label{Toddasym}
\eeq
where $\Gamma$ is the decay rate for the process in question.

The most obvious place to search for triple products is in the decay
$B\to V_1 V_2$, where both $V_1$ and $V_2$ are vector mesons. In this
case, the TP takes the form ${\vec\varepsilon}_1^{*\sss T} \times
{\vec\varepsilon}_2^{*\sss T} \cdot {\hat p}$, where ${\vec p}$ is the
momentum of one of the final vector mesons, and ${\vec\varepsilon}_1$
and ${\vec\varepsilon}_2$ are the polarizations of $V_1$ and $V_2$.
Note that TP's can be obtained by performing an angular analysis of
the $B \to V_1 V_2$ decay. However, as seen from Eq.~(\ref{Toddasym}),
a full angular analysis is not necessary.

Now, because triple products are odd under T, they can be faked by
strong phases. That is, one can obtain a TP signal even if the weak
phases are zero. In order to obtain a true CP-violating signal, one
has to compare the TP in $B \to V_1 V_2$ with that in ${\bar B} \to
{\bar V}_1 {\bar V}_2$. The CP-violating TP is found by {\it adding}
the two T-odd asymmetries \cite{Valencia}:
\beq
{\cal A}_{\sss T} \equiv {1\over 2}(A_{\sss T} + {\bar A}_{\sss T}) ~.
\label{CPTP}
\eeq
Thus, neither tagging nor time dependence is necessary to measure
TP's. One can in principle combine measurements of charged and neutral
$B$ decays \cite{LSS}.

The main point is that the CP-violating TP asymmetry of
Eq.~(\ref{CPTP}) takes the form
\beq
{\cal A}_{\sss T} \propto \sin\phi \cos\delta ~.
\label{TPsignal}
\eeq
That is, unlike ${\cal A}_{\sss CP}^{dir}$ [Eq.~(\ref{directCPV})],
the triple product does not vanish if $\delta = 0$. Thus, TP's are
complementary to direct CP asymmetries. In order to completely test
for the presence of NP, it is necessary to measure both direct CP
violation and triple products.

This then begs the question: which $B\to V_1 V_2$ decays are expected
to yield large TP's in the SM? Interestingly, the answer is {\it none}
\cite{Valencia,BVVTPs,AtwoodSoni,DatLonTP}! It is straightforward to
see how this comes about.

As noted above, all CP-violating effects require the interference of
two amplitudes, with different weak phases. Thus, there can be no
triple products in decays which in the SM are dominated by a single
decay amplitude.

Now consider other $B\to V_1 V_2$ decays. Within factorization, the
amplitude can be written
\beq
\sum_{{\cal O},{\cal O}'} \left\{ \bra{V_1}
{\cal O} \ket{0} \bra{V_2} {\cal O}' \ket{B} + \bra{V_2} {\cal O}
\ket{0} \bra{V_1} {\cal O}' \ket{B} \right\} ~,
\label{TPops}
\eeq
where ${\cal O}$ and ${\cal O}'$ are SM operators. The key point is
that TP's are a {\it kinematical} CP-violating effect \cite{Kayser}.
That is, in order to produce a TP in a given decay, both of the above
amplitudes must be present, with a relative weak phase.

For example, consider the decay $\bd \to D^{*+} D^{*-}$. There is a
tree amplitude, proportional to $V_{cb}^* V_{cd}$, and a penguin
amplitude, proportional to $V_{tb}^* V_{td}$. Given that there are two
amplitudes with a relative weak phase, one would guess that a
CP-violating triple product would be produced. However, this is not
the case. In fact, both amplitudes contribute to the $\bra{D^{*+}}
{\cal O} \ket{0} \bra{D^{*-}} {\cal O}' \ket{B}$ matrix elements; there
is no $\bra{D^{*-}} {\cal O} \ket{0} \bra{D^{*+}} {\cal O}'
\ket{B}$. (That is, in the SM one has only ${\bar b} \to {\bar c}$
transitions; ${\bar b} \to c$ transitions do not occur.) Thus, despite
the presence of two amplitudes in this decay, no TP is produced, at
least within factorization.

Using the above argument, we note that there are three classes of
$B\to V_1 V_2$ decays in the SM, all of which are expected to have
zero or small triple products:
\begin{enumerate}

\item Decays governed by a single weak decay amplitude, such as $B \to
J/\psi K^*$, $\bs \to \phi \phi$, $\bs \to D_s^* D_s^*$, $B_c^+ \to
J/\psi \rho^+$, etc. Because there is only one amplitude, there can be
no CP-violating effects, including TP's. This is model-independent.

\item Color-allowed decays with two weak decay amplitudes, such as
$\bdbar \to D^{*+} D^{*-}$, $\bsbar \to D_s^{*+} D^{*-}$, $\bsbar \to
K^{*+} K^{*-}$, $B_c^- \to {\bar D}^{*0} \rho^-$, $B_c^- \to {\bar
D}^{*0} K^{*-}$, etc. The two amplitudes are usually a tree and a
penguin diagram, though it is possible to have two penguin
contributions. As argued above, both decay amplitudes contribute to
the same kinematical amplitude, so that all TP's vanish. Since the
decays are colour-allowed, nonfactorizable corrections are expected to
be small, so that the prediction of tiny TP's is robust.

\item Color-suppressed decays with two weak amplitudes, such as
$B^-\to \rho^0 K^{*-}$, $\bsbar\to \phi K^{*}$, $B_c^-\to J/\psi
D^{*-}$, etc. Once again, within factorization, both amplitudes
contribute to the same kinematical amplitude, so that the TP's
vanish. However, nonfactorizable effects may be large in
colour-suppressed decays. We have tried to be conservative in our
estimates of such effects, and still find tiny TP's for such decays
\cite{DatLonTP}. This conclusion is clearly model-dependent. (In any
case, the branching ratios for such decays are very small.)

\end{enumerate}

The fact that all TP's in $B\to V_1 V_2$ are expected to vanish or be
very small in the SM makes this an excellent class of measurements to
search for new physics. In the SM, all couplings to the $b$-quark
[i.e.\ the operators ${\cal O}'$ in Eq.~(\ref{TPops})] are
left-handed. Within factorization, the discovery of a large TP in a
$B\to V_1 V_2$ decay would point to new physics with large couplings
to the right-handed $b$-quark \cite{AtwoodSoni,DatLonTP}. Many
new-physics models, though not all, have such couplings. As an
example, supersymmetry with R-parity-violating couplings can explain
the apparent discrepancy in $\bd(t) \to \phi \ks$. Such a model of NP
will also contribute to $B \to \phi K^*$ decays, leading to TP's. In
the SM, such TP's vanish; with this type of NP, one can get TP
asymmetries as large as 15--20\% \cite{DatLonTP}! The upshot is that
triple products are excellent diagnostic tests for new physics. Some
NP models predict large TP's, so that null measurements can strongly
constrain (or eliminate) such models.

\section{Time-Dependent Angular Analysis}

Consider now a $V_1 V_2$ state which in the SM is dominated by a
single amplitude. Suppose that there is a new-physics amplitude, with
a different weak phase, contributing to this decay. Above, I have
argued that one can detect such an amplitude by looking for both
direct CP violation and triple products. However, much more
information can be obtained if a time-dependent angular analysis of
the corresponding $B^0(t) \to V_1 V_2$ decay can be performed
\cite{LSS}.

The time-dependent decay rate for $B^0(t) \to V_1 V_2$ is given by
\beq
\Gamma(B^0(t) \to V_1 V_2) = e^{-\Gamma t}
\sum_{\lambda\leq\sigma} \left( \Lambda_{\lambda\sigma} +
\Sigma_{\lambda\sigma}\cos(\Delta M t) 
- \rho_{\lambda\sigma}\sin(\Delta M t) \right) 
g_\lambda g_\sigma ~.
\eeq
In the above, the helicity indices $\lambda$ and $\sigma$ take the
values $\left\{0,\|,\perp \right\}$, and the $g_\lambda$ are known
functions of the kinematic angles. For a given helicity $\lambda$,
$\Lambda_{\lambda\lambda}$ essentially measures the total rate, while
$\Sigma_{\lambda\lambda}$ and $\rho_{\lambda\lambda}$ represent the
direct and indirect CP asymmetries, respectively. The quantity
$\Lambda_{\perp i}$ ($i=\{0,\|\}$) is simply the triple product
discussed earlier.

Now, there are 18 observables in this decay rate. However, if there is
no new physics, there are only 6 theoretical parameters. This implies
that there are 12 relations among the observables. They are:
\bea
& \Sigma_{\lambda\lambda}= \Lambda_{\perp i}= \Sigma_{\| 0}=0 ~, & \nn\\
& \rho_{ii} / \Lambda_{ii} = - \rho_{\perp\perp} / \Lambda_{\perp\perp}
= \rho_{\|0} / \Lambda_{\| 0} ~, & \nn\\
& 2 \Lambda_{\|0} \Lambda_{\perp\perp} \left(
\Lambda_{\lambda\lambda}^2 -\rho_{\lambda\lambda}^2 \right) = 
\left[ \Lambda_{\lambda\lambda}^2 \rho_{\perp 0} \rho_{\perp\|} +
  \Sigma_{\perp 0} \Sigma_{\perp \|} \left( \Lambda_{\lambda\lambda}^2
  -\rho_{\lambda\lambda}^2 \right) \right] ~, & \nn\\
& \rho_{\perp i}^2 \Lambda_{\perp\perp}^2 = \left(
\Lambda_{\perp\perp}^2 -\rho_{\perp\perp}^2 \right) \left(
4\Lambda_{\perp\perp} \Lambda_{ii}-\Sigma_{\perp i}^2 \right) ~. &
\eea
The violation of any of these relations will be a smoking-gun signal
of NP. If the NP conspires to make direct CP violation and TP's small,
it can still be detected through one of these other signals
\cite{LSS}. Thus, if a time-dependent angular analysis can be
performed, there are many more ways to search for new physics.

However, even more can be done! Suppose that some signal for new
physics is found. In this case, it is straightforward to show that
there are more theoretical parameters than independent observables, so
that one cannot solve for the NP parameters. However, because the
expressions relating the observables to the theoretical parameters are
nonlinear, one can actually put a {\it lower bound} on the NP
parameters \cite{LSS}. This is extremely important, as it allows us to
get direct information on the NP through measurements in the $B$
system.

\section{$\alpha$ from $B^0 \to K^{(*)} {\bar K}^{(*)}$ Decays}

I now turn to the extraction of $\alpha$ from $B^0_{d,s} \to K^{(*)}
{\bar K}^{(*)}$ decays \cite{DatLonalpha}. Consider first $\bd\to K^0
\kbar$, which is a pure $b\to d$ penguin:
\bea
A(\bd\to K^0 \kbar) & = & P_u \, V_{ub}^* V_{ud} + P_c \, V_{cb}^*
V_{cd} + P_t \, V_{tb}^* V_{td} \nn\\
& = & \Puc \, e^{i\gamma} \, e^{i\delta_{uc}} + \Ptc \, e^{-i\beta} \,
e^{i\delta_{tc}} ~,
\eea
where $\Puc\equiv |(P_u - P_c) V_{ub}^* V_{ud}|$, $\Ptc\equiv |(P_t -
P_c) V_{tb}^* V_{td}|$, and I have explicitly written out the strong
phases $\delta_{uc}$ and $\delta_{tc}$, as well as the weak phases
$\beta$ and $\gamma$. By measuring $\bd(t) \to K^0 \kbar$, one can
extract 3 observables -- the total rate, and the direct and indirect
CP asymmetries. However, these depend on the 4 unknowns $\Puc$,
$\Ptc$, $\Delta \equiv \delta_{uc} - \delta_{tc}$ and $\alpha$, so
that CP phase information cannot be obtained.

Now consider a second pure $b\to d$ penguin decay of the form $\bd \to
K^* \kstarbar$, where $K^*$ represents any excited neutral kaon. This
decay can be treated completely analogously to $\bd\to K^0 \kbar$,
with unprimed parameters and observables being replaced by ones with
tildes. The measurement of the time-dependent rate again allows one to
extract 3 observables, which depend on the 4 unknowns $\tildePuc$,
$\tildePtc$, $\tildeDelta$ and $\alpha$. Again, there are more
observables than unknowns, so that one cannot extract $\alpha$.
However, one can combine measurements from the two decays to write
\beq
{\Ptc^2 \over {\tildePtc}^2} = f(\alpha,{\rm observables}) ~.
\label{alphasolve}
\eeq
Note that the CKM matrix elements $|V_{tb}^* V_{td}|$ cancel in this
ratio. From this ratio, we see that we could solve for $\alpha$ if we
knew the value of $\Ptc/\tildePtc$.

This information can be obtained by considering $\bs \to K^{(*)} {\bar
K}^{(*)}$ decays. Consider the decay $\bs\to K^0\kbar$, which is a
pure $b\to s$ penguin:
\bea
A(\bs\to K^0\kbar) & = & P_u' \, V_{ub}^* V_{us} + P_c' \, V_{cb}^*
V_{cs} + P_t' \, V_{tb}^* V_{ts} \nn\\
& = & \Puc' \, e^{i\gamma} \, e^{i\delta_{uc}'} + \Ptc' \,
e^{i\delta_{tc}'} \nn\\ 
& \simeq & \Ptc' \, e^{i\delta_{tc}'} ~.
\eea
Here $\Puc' \equiv |(P_u' - P_c') V_{ub}^* V_{us}|$ and $\Ptc' \equiv
|(P_t' - P_c') V_{tb}^* V_{ts}|$. However, $\left\vert {V_{ub}^*
  V_{us} / V_{tb}^* V_{ts}} \right\vert \simeq 2\%$, so that the
$u$-quark piece $\Puc'$ is negligible compared to $\Ptc'$, leading to
the last line above. Therefore the measurement of $B(\bs\to K^0\kbar)$
gives $\Ptc'$. Similarly, the measurement of $B(\bs \to K^*
\kstarbar)$ gives $\tildePtc'$.

The key point is that, in the flavour SU(3) limit, we have
$\Ptc'/\tildePtc' = \Ptc/\tildePtc$. Thus, using
Eq.~(\ref{alphasolve}), one can obtain $\alpha$. Now, uncertainties
due to SU(3) breaking are typically $\sim 25\%$. However, these
leading-order effects {\it cancel} in the double ratio
$(\Ptc'/\tildePtc')/(\Ptc/\tildePtc)$. One is left with only
second-order SU(3)-breaking effects, i.e.\ a theoretical error of at
most 5\% \cite{DatLonalpha}.

One can compare the value of $\alpha$ extracted with this method with
that obtained elsewhere (e.g.\ in $B\to\pi\pi$ or $B\to\rho\pi$). A
discrepancy would point to new physics in the $b\to d$ or $b\to s$
penguin.

It is useful to list some experimental considerations. First, $\bs$
mesons are involved, and all branching ratios are $\sim 10^{-6}$.
Second, the $K^{(*)}$ and ${\bar K}^{(*)}$ mesons are detected through
their decays to charged $\pi$'s and $K$'s only, requiring good $K/\pi$
separation. Finally, no $\pi^0$ detection needed. All in all, this
method is particularly appropriate to hadron colliders.

\section{Triple products in $\Lambda_b$ Decays}

Another class of decays which can only be studied at hadron colliders
are those involving $\Lambda_b$ baryons. Consider the decays
$\Lambda_b \to F_1 P$ and $F_1 V$, where $F_1$ is a fermion ($p$,
$\Lambda$, ...), $P$ is a pseudoscalar ($K^-$, $\eta$, ...), and $V$
is a vector ($K^{*-}$, $\phi$, ...). In these decays, triple products
are possible \cite{wafia1}. In $\Lambda_b \to F_1 P$, only one TP is
possible: $\vec p_{F_1} \cdot (\vec s_{F_1} \times \vec
s_{\Lambda_b})$, where ${\vec s}_i$ is the spin of particle $i$. On
the other hand, since the decay $\Lambda_b \to F_1 V$ involves three
spins and one final momentum, four TP's are possible.

As in $B\to V_1 V_2$ decays, within factorization we require a
right-handed coupling to $b$-quarks in order to generate a TP. For
certain $F_1 P$ final states, one can ``grow'' a sizeable right-handed
current due to the Fierz transformations of some of the SM operators.
However, for $F_1 V$ final states, there are no such right-handed
currents. Thus, all TP's are expected to vanish in the SM for
$\Lambda_b \to F_1 V$ decays.

We find that ${\cal A}_T^{pK} = -18\%$, but the TP's for all other
fermion-pseudoscalar final states ($p K^{*-}$, $\Lambda \eta$,
$\Lambda\eta'$, $\Lambda\phi$) are small, at most $O(1\%)$. Once
again, the fact that almost all TP's are expected to be small implies
that this is a good place to look for new physics. In fact, one can
use TP's in $\Lambda_b$ decays as a diagnostic tool for NP
\cite{wafia2}.

\section{Radiative Decays}

Finally, I consider radiative decays of $B$ mesons. The inclusive
partial rate asymmetries can be calculated reliably in the SM
\cite{Hurth}:
\bea
{\cal A}_{CP}^{dir} (b \to s\gamma) & = & 0.5\% ~, \nn\\
{\cal A}_{CP}^{dir} (b \to d\gamma) & = & -10\% ~.
\eea
If measurements of these asymmetries are found to differ from their SM
values, this will indicate the presence of new physics. Indeed, large
deviations are possible in several models of NP \cite{radNP}.

Exclusive partial rate asymmetries in $B \to K^* \gamma$ and $B \to
\rho\gamma$ are not known as well -- there are important bound-state
corrections \cite{GSW}. However, if significant deviations from the
values calculated for inclusive decays are found in exclusive decays,
this probably points to new physics.

One can also consider mixing-induced CP asymmetries (e.g.\
$\bd(t)\to\rho\gamma$, $\bs(t) \to \phi\gamma$). In the SM the photon
polarization is opposite for $B$ and ${\bar B}$ decays, so that no
interference is possible. That is, ${\cal A}_{CP}^{mix} (b \to
s\gamma, b\to d\gamma) \simeq 0$ in the SM. However, one can get a
significant ${\cal A}_{CP}^{mix}$ in certain models of NP (e.g.\
left-right symmetric models, SUSY, models with exotic fermions)
\cite{AGS}.

\section{Conclusion}

In summary, there are numerous signals of new physics in $B$ and
$\Lambda_b$ decays. Furthermore, there are many ways of determining
which types of NP might be responsible for these signals. If the NP is
discovered directly, the measurement of CP violation in the $B$ system
can be used to probe its couplings. Thus, the study of $B$ processes
is complementary to direct searches for NP. Hadron colliders have a
significant role to play in the discovery of NP, as well as in its
identification.

\bigskip
\noindent
{\bf Acknowledgements}:
%\bigskip
This work was financially supported by NSERC of Canada.

\end{document}